\documentclass{Interspeech}
\usepackage{booktabs}
\usepackage{tabularx}
\usepackage{multirow}
\usepackage{siunitx}
\usepackage{array}
\interspeechfinaltrue

\title{Inter-Speaker Relative Cues for Text-Guided Target Speech Extraction}

\author[affiliation={}]{Wang}{Dai}
\author[affiliation={}]{Archontis}{Politis}
\author[affiliation={}]{Tuomas}{Virtanen}

\affiliation{Audio Research Group}{Tampere University}{Finland}
\email{\texttt{\{name.surname\}@tuni.fi}}


\usepackage{comment}

\begin{document}

\maketitle

\begin{abstract}
    
We propose a novel approach that utilizes inter-speaker relative cues to distinguish target speakers and extract their voices from mixtures. Continuous cues (e.g., temporal order, age, pitch level) are grouped by relative differences, while discrete cues (e.g., language, gender, emotion) retain their categorical distinctions. Compared to fixed speech attribute classification, inter-speaker relative cues offer greater flexibility, facilitating much easier expansion of text-guided target speech extraction datasets. Our experiments show that combining all relative cues yields better performance than random subsets, with gender and temporal order being the most robust across languages and reverberant conditions. Additional cues, such as pitch level, loudness, distance, speaking duration, language, and pitch range, also demonstrate notable benefits in complex scenarios. Fine-tuning pre-trained WavLM Base+ CNN encoders improves overall performance over the Conv1d baseline.
\end{abstract}

\section{Introduction}

Target speech extraction (TSE) methods aim at isolating a specific speaker’s voice from a multi-talker mixture. This process uses cues associated with the desired speaker, such as a pre-recorded enrollment speech that highlights the speaker’s vocal characteristics \cite{Wang2019, vzmolikova2019speakerbeam}, a spatial cue indicating the direction from which the speaker is speaking \cite{ge2022spex}, or video input capturing lip movements \cite{ochiai2019multimodal, pan2022selective}. However, in real-world scenarios, these pre-registered cues can vary significantly, raise privacy concerns or may even be absent, thereby limiting the practicality and effectiveness of TSE systems.

Compared to the aforementioned cues, text is more accessible, making it a practical and flexible option. With advances in language models, several studies have used natural language descriptions to extract specific sound events or musical instruments from audio mixtures, achieving impressive results \cite{liu2024separate, li2023target, jiang2024listen}. Building on this, recent study has employed text descriptions to isolate a specific speaker's voice from a speech mixture. LLM-TSE \cite{hao2023typing} was the first to use natural language descriptions powered by a large language model to extract semantic information from text queries (such as a description of single speech attribute, e.g., gender, language) and integrate it into a speech separation network, demonstrating the effectiveness of text-guided TSE.

However, its focus on a single speech attribute limits its broader application. To overcome this, Huo et al. introduced a more flexible text-guided TSE model and developed the TextrolMix dataset, which provides rich descriptions of speech attributes \cite{10890201}. This dataset offers two types of contextual clues: (1) a natural language description of the speaking style of the target speech and (2) a reference audio that shares specific style attributes with the target speech, accompanied by a corresponding text prompt. Each utterance is annotated with six attributes: speaker identity, emotion, pitch level, gender, accent, and tempo (speaking rate). The speaker identity is represented by an enrollment speech sample from the target speaker, aligned with the relevant text prompt.

Despite these advancements, certain limitations persist. Firstly, the previous studies categorize continuous-valued attributes, such as pitch level, based on single-speaker speech statistics \cite{jiang2024listen, hao2023typing, 10890201}, which results in coarse quantization and the loss of valuable information. Secondly, the TextrolMix dataset captures only a limited set of speech attributes, omitting factors like language, temporal order, and speaking duration. Furthermore, it does not guarantee that the mixture contains different speakers, limiting its robustness in real-world scenarios. Similarly, Jiang et al. \cite{jiang2024listen} use an LLM to interpret generic text prompts (queries) for target speech extraction with TextrolSpeech dataset \cite{ji2024textrolspeech}, but encounter the same issues, especially the narrow scope of speech attributes and inflexible categorization.

To address these limitations, we incorporate a broad range of speech attributes and create a two-speaker mixture dataset, including language, gender, transcription, emotion, temporal order, age, speaking rate, duration, pitch level (mean $F_0$), pitch range ($F_0$ span), loudness, and speaker-to-microphone distance. The contributions of our work are as follows: (i) We introduce inter-speaker relative cues, which categorize the collected attributes based on the differences between the target and interference speakers' voices. The relative cues provide a way to overcome the fixed categorization of speech attributes and make the dataset much easier to expand. (ii) We conduct baseline experiment to evaluate the impact and contribution of these cues in extracting target speakers' voices. (iii) Since the addition of more cues can complicate the learning process, we propose utilizing pre-trained weights of WavLM Base+ CNN encoders for improved extraction performance. 
To facilitate future research in this direction, code for creating our dataset is available at: \url{https://github.com/daiwangsnr}.

\section{Dataset Construction}
In this study, mixture signals consist of speech from two speakers: one as the target and the other as interference. We construct relative cues from various speech attributes to specify and extract the target speaker's voice. These cues are formatted as natural language descriptions (text prompts) for target speech extraction. Our dataset includes three components: inter-speaker relative cues, audio part, and text prompt generation.

\subsection{Inter-Speaker Relative Cues}

Inter-speaker relative cues are constructed based on multiple speech attributes and applied to each pair of target and interference speech in one mixture.
We first compile metadata for both the target and interference speech samples, covering attributes such as language, gender, age, emotional state, transcription, mean $F_0$, $F_0$ span, signal-to-interference Ratio (SIR), speaker-to-microphone distance, appearance time in the mixture, speaking duration, and speaking rate. 

The mean $F_0$ and $F_0$ span are extracted using the Librosa pYIN library\footnote{\url{https://librosa.org/doc/main/generated/librosa.pyin.html}}, with the $F_0$ span measured logarithmically to reduce the impact of outliers. 
If transcription annotations are available, word and pause durations are estimated using word time boundaries from the Montreal Forced Aligner tool\footnote{\url{https://montreal-forced-aligner.readthedocs.io/en/latest/}}. Otherwise, we use the Silero VAD tool\footnote{\url{https://github.com/snakers4/silero-vad}} to detect word and pause durations. The speaking duration is calculated by summing the durations of words and brief pauses, while excluding pauses longer than 0.6 seconds. Pauses shorter than this threshold are considered natural breaks.
The speaking rate is calculated by dividing the total number of syllables by the speaking duration. Syllable counting is based on the number of vowel sequences, where in Chinese each character (acquired by splitting the text into individual characters) is counted as a syllable , while in English, French, German, and Spanish, each syllable is estimated by detecting continuous vowel groups (acquired by splitting words into tokens and identifying consecutive vowel letters such as ``aeiou''). 
The SIR is calculated with consideration of the speaking durations of both speech samples to estimate their relative loudness.
The speaker-to-microphone distance for each speech sample is generated by our random constraining of speaker in a virtual room (for further details, refer to Section 2.2). Language, gender, age, emotional state, and transcription are derived from the original annotation information in the datasets used (refer to Section 2.2). 

Relative cues are derived by comparing target and interference speech attributes. For continuous-valued attributes (e.g., age, mean $F_0$, $F_0$ span, speaking duration, speaking rate, SIR, distance), cues are categorized based on auditory thresholds (see the bold numbers in Table 1). If the attribute difference exceeds the threshold, cues are assigned to two separate groups (e.g., older/younger for age category cue), otherwise they are labeled as ``similar''. For example, if speaking rates differ by more than 15\%, the cue is labeled as ``faster'' or ``slower''; otherwise, it is labeled ``similar''. For discrete attributes (e.g., language, transcription, gender, emotion), cues are labeled as ``same'' if attribute values match; otherwise, the specific attribute is used (e.g., English/French for language, spoken words transcription for transcription, male/female for gender, happy/angry for emotion). The threshold for temporal order is based on our subjective listening test, while other thresholds are derived or refined from hearing studies on speech signals, considering cross-language variations and reverberant conditions.  
\begin{table}[h]
    \centering
    \caption{Used auditory thresholds for some relative speech cues}
    \label{tab:perceptual_thresholds}
    \renewcommand{\arraystretch}{1.0}
    \begin{tabular}{p{3.2cm}|p{3.8cm}} 
        \hline
        \textbf{Cue} & \textbf{Threshold} \\
        \hline
        Speaking Rate & \textbf{15\%} \\
        (Faster/Slower/Similar) & 5-10\% noticeable \cite{miller1981some, quene2007just} \\
        \hline
        Speaking Duration & \textbf{15\%} \\
        (Longer/Shorter/Similar) & 5-10\% noticeable \cite{miller1981some} \\
        \hline
        Pitch Level & \textbf{5 Hz} \\
        (Higher/Lower/Similar) & 3-5 Hz perceptible \cite{moore2012introduction} \\
        \hline
        Pitch Range & \textbf{25\%} \\
        (Wider/Narrower/Similar) & 10-20\% perceptible \cite{moore2012introduction} \\
        \hline
        Distance & \textbf{0.5 m} \\
        (Farther/Nearer/Similar) & 20-30 cm noticeable \cite{blauert1997psychophysics} \\
        \hline
        Age Category & \textbf{10 years} \\
        (Older/Younger/Similar) & 5-10 years discernible \cite{linville2001vocal, ryan1978age} \\
        \hline
        Loudness & \textbf{3 dB} \\
        (Louder/Quieter/Similar) & 3 dB noticeable \cite{blauert1997psychophysics, zwicker2013psychoacoustics} \\
        \hline
        Temporal Order & \textbf{0.1 s} \\
        (First/Second/Similar) & 0.1 s noticeable in our subjective listening test  \\
        \hline
    \end{tabular}
\end{table}

\subsection{Audio part}
\begin{table}[h]
\centering
\caption{Summary of used speech datasets across five languages, detailing the number of speakers in the training, validation, and test splits, along with their partial attributes.}
\label{tab:emotion_datasets}
\renewcommand{\arraystretch}{1.0}
\setlength{\tabcolsep}{4.5pt}
\begin{tabular}{p{2.22cm}|p{1.15cm}|p{1.1cm}|p{2.3cm}}
\hline
\textbf{Dataset} & \textbf{Language} & \textbf{Speakers} & \textbf{Attributes} \\ \hline
CASIA \cite{zhang2008design}        & Chinese            & 2/1/1                    & Language, Emotion, Gender, Transcription \\ \hline
ESD \cite{zhou2022emotional}            & Chinese            & 6/2/2              & Language, Emotion, Gender, Transcription \\ \hline
Aishell3 \cite{shi21c_interspeech}      & Chinese          & 120/30/30              & Language, Gender, Transcription \\ \hline
MagicData-Read \cite{magicdata2019}         & Chinese          & 33/8/11                   & Language, Age, Gender \\ \hline
Librispeech \cite{panayotov2015librispeech}    & English        & 120/30/30   & Language, Gender, Transcription           \\ \hline
ESD \cite{zhou2022emotional}            & English           & 6/2/2               & Language, Emotion, Gender, Transcription \\ \hline
OreauFR-02 \cite{kerkeni2020french}    & French         & 21/7/4                      & Language, Emotion, Age, Gender \\ \hline
MLS French \cite{Pratap2020MLSAL}    & French         & 90/18/18                    & Language, Gender, Transcription \\ \hline
EmoDB \cite{burkhardt2005database}         & German          & 6/2/2                     & Language, Emotion, Age, Gender, Transcription \\ \hline
MLS German \cite{Pratap2020MLSAL}     & German         & 120/30/30                     & Language, Gender, Transcription \\ \hline
MLS Spanish \cite{Pratap2020MLSAL}   & Spanish       & 79/20/20                    & Language, Gender, Transcription \\ \hline
EmoMatchSpanish \cite{garcia2024emomatchspanishdb} & Spanish       & 30/10/10         & Language, Emotion, Gender, Transcription \\ \hline
\end{tabular}
\end{table}
Our goal is to create a mixed speech dataset with diverse inter-speaker relative cues (attributes). Since no single corpus provides all required attributes, we combine speech from multiple corpora, selecting representative datasets across five major languages (summarized in Table 2). These corpora include rich attributes like language, gender, transcription, age, and emotional state, enabling the generation of diverse speech mixtures. 
The speech mixture creation process involves the following steps:

(i) To avoid overlaps in speakers and spoken content across training, validation, and test sets, and to expedite mixing, we split each corpus into two non-overlapping parts, ensuring no shared speakers or content. These parts are merged into two larger sets: training1/validation1/test1 and training2/validation2/test2. The Montreal Forced Aligner tool is used to obtain word time boundaries for utterances with transcriptions.

(ii) We generate 100,000 mixed speech samples for training, and 10,000 each for validation and testing. Due to limited emotion and age annotations, we create three sub-pools: one with emotion labels, one with age labels, and one without either. We sample pairs from these sub-pools as follows: 20,000 training, 2,000 validation, and 2,000 test pairs for the emotion sub-pool; 10,000 training, 1,000 validation, and 1,000 test pairs for the age sub-pool; and 70,000 training, 7,000 validation, and 7,000 test pairs for the third sub-pool.
During each mixing iteration, two speech samples are randomly selected from the sub-pools. Speech durations are capped at 6 seconds; longer utterances are truncated, and shorter ones are retained. Silence at the beginning or end is removed using word alignment (for transcribed speech) or voice activity detection. The trimmed samples $S_1$ and $S_2$ retain their original start and end times for transcription cues.

(iii) In two-talker speech, there is usually a temporal order, with one speaker speaking first and the other following, often with some overlap. If either $S_1$ or $S_2$ is shorter than 3 seconds, the overlap equals the duration of the shorter sample, padded with zeros at the beginning using a randomly selected start offset from \( (0, \text{duration of longer sample} - \text{duration of shorter sample}) \).
For samples longer than 3 seconds, the overlap is the difference between their total durations and 6 seconds, with $S_2$ padded at a fixed offset. The padded versions, $S_1'$ and $S_2'$ retain word boundaries from the original transcriptions, enabling transcription cues. Either $S_1'$ or $S_2'$ serves as the target clean speech.

(iv) We simulate reverberant speech using the generated clean speech samples and gpuRIR library \cite{diaz2021gpurir}, which models speakers in a room with reverberation effects. The room dimensions range from 9–11 m in length and width, 2.6–3.5 m in height, and reverberation times of 0.3–0.6 seconds.
The microphone is positioned at the center of the room, while speaker-to-microphone distances vary from 0.3 to 1.5 m horizontally, with speaker heights between 1.6 and 1.9 m.
We generate 10,000 RIR pairs for training, 1,000 for validation, and 1,000 for testing. For each mixing iteration, we randomly select RIRs ($rir_1$, $rir_2$) and convolve them with clean speech signals ($S_1'$, $S_2'$). The SIR between each reverberant speech pair is uniformly sampled from -6 to 6 dB before merging them into a single mixture.

\subsection{Text Prompt Generation}
The goal is to create generic text prompts that capture inter-speaker relative cues for target speech extraction. We start by compiling template prompts based on one or more relative cues, following a similar approach to \cite{jiang2024listen}. Since individual relative cues are intended to specify the target speaker, cues classified as ``same" or ``similar'' are excluded. The template prompt is generated by inserting actions and cues into structures like \texttt{"Please <verb> <str(cues)>."} or \texttt{"Can you <verb> str(cues)?"}. Here, the \texttt{verb} represents an action, randomly selected from ``extract", ``isolate", and ``separate", while the \texttt{str} function converts a cue vector into a human-readable string by inserting a noun phrase, such as ``speaker characterized by". For instance, the cue vector (female, higher pitch level, faster speaking rate) can be expressed as: ``the female speaker characterized by a higher pitch level and a faster speaking rate".

For each mixed speech sample in the TSE training, validation, and test sets, we generate multiple template prompts, including twelve individual cues, a randomly selected subset of cues (random cues), and a combination of all individual cues (all cues). To enhance template flexibility, we rephrase each prompt into five variations using ChatGPT-4o-mini, leveraging its efficiency and accuracy. We instruct the model to focus on action verbs such as ``extract'', ``separate'' and ``isolate'' while avoiding terms like ``identify'' and ``locate''. In addition, we encourage the model to explore different sentence structures and synonyms for keywords that describe speech or speaker characteristics, such as ``female'', ``male'', ``pitch range'', ``speaking speed'', ``quieter loudness'', ``louder loudness'', ``longer speaking duration'', ``shorter speaking duration'', and so on. Furthermore, we instruct the model to ensure the language remains simple and suitable for everyday use.
After generating these generic prompts, we found that only a few outputs were incomplete or irrelevant. These were manually removed and regenerated. 

\section{Experimental Setup}

\subsection{Proposed Method and Baseline}
The architecture of our proposed method is depicted in Figure 1. We adopt the conventional encoder-mask-decoder framework for speech separation, where the text prompt, describing cues of the target speaker, guides the extraction of the target speaker from the mixture speech.
\begin{figure*}[t]
  \centering
  \includegraphics[width=0.96\linewidth, height=4.8cm]{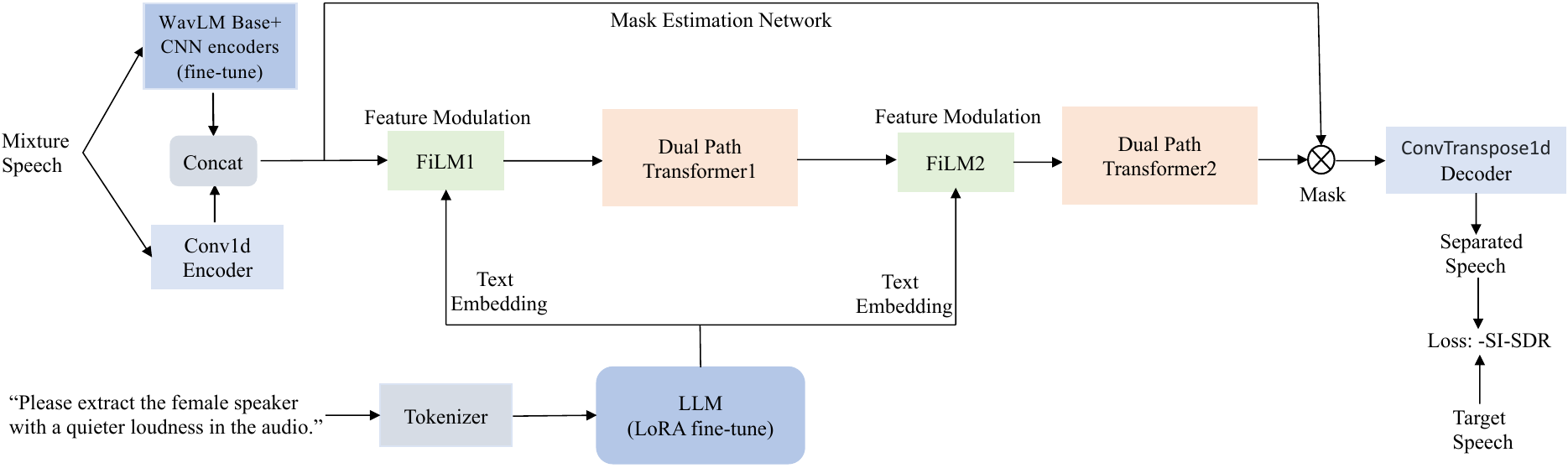}
  \caption{A block diagram of the used target speech extraction system}
\end{figure*}

To enhance learning of mixed speech signals and interpret target speech cues, we leverage self-supervised representations pre-trained on overlapping speech data and use a large language model to encode these cues into a unified embedding space. Specifically, we incorporate CNN encoders from the WavLM Base+ model as an additional encoder for the mixture speech. WavLM Base+ is pre-trained with masked speech denoising and prediction techniques, learning from masked segments to predict pseudo-labels. This improves downstream tasks like speech recognition, speaker verification, separation, and diarization. By training on noisy, overlapped speech, its CNN encoders capture rich speech and speaker features, which have been proven to benefit target speech extraction \cite{peng2024target}. We fine-tune the WavLM Base+ CNN encoders and concatenate their output with the Conv1d encoder along the feature dimension.

To model both local and global speech dependencies, we use two dual-path transformer (DPT) blocks \cite{subakan2021attention}, each with four intra- and four inter-transformer layers for mask estimation. To better interpret text cues, we employ an LLM, specifically the resource-efficient LLaMA 3.2 1B Instruct model, leveraging its strong multilingual understanding and instruction-following capabilities. The network incorporates text embeddings from the fine-tuned LLaMA 3.2 1B Instruct model. These embeddings are fused with the encoder output and the first DPT block output via FiLM modules, similar to \cite{jiang2024listen}. The language model is jointly trained with the mask estimation network to generate text embeddings that exhibit awareness of relative cues written in the text prompt, facilitating the system identify the target speaker.
To assess the benefits of additional pre-trained features, we introduce a baseline model identical to the proposed one but without WavLM Base+ CNN encoders.

\subsection{Training and Network Configurations}
We adopted a relative large kernel size of 80 and stride of 40 for both the Conv1d encoder and ConvTranspose1d decoder, similar to \cite{jiang2024listen}, which proved efficient in preliminary experiments. To match the time resolution of the Conv1d encoder, only the first four temporal convolution blocks of WavLM Base+ CNN encoders were fully fine-tuned and used for feature concatenation. Intra- and inter-transformers processed 50-frame chunks with 50\% overlap, using 8 attention heads and 2048-dimensional feed-forward networks per layer.

The LLaMA 3.2 1B instruction model was fine-tuned using LoRA (rank 16, scaling factor 16, dropout 0.05), applied to query and key projection layers in self-attention. Training was performed using the negative SI-SDR \cite{le2019sdr} loss with float16 mixed-precision and a batch size of 24. The AdamW optimizer was used with a learning rate of 1e-4, halved if validation loss did not decrease for three consecutive epochs. 
A linear warm-up was applied during the first 1,000 updates, and gradient normalization (with a maximum norm of 30) was employed to prevent gradient explosion. Models were trained with early stopping, with up to 100 epochs.
Following \cite{jiang2024listen}, we used all five generic text prompts with a randomly selected subset of cues (random cues) for each mixture during training. For validation, only the fifth set of generic prompts with random cues were used.

\subsection{Results and Discussion}
We evaluate target speech extraction performance using SI-SDR improvement (SI-SDRi) and perceptual evaluation of speech quality (PESQ) \cite{941023}. The fifth set of generic prompts in the test set, which includes random, individual, and all cues for each test sample, is used for evaluation. Table 3 presents the results when both models are trained with random cues.

\begin{table}[h]
\centering
\caption{Performance of baseline and proposed method}
\setlength{\tabcolsep}{5pt}  
\begin{tabular}{>{\raggedright\arraybackslash}p{2.4cm}|S[table-format=1.]|S[table-format=1.1]|S[table-format=1.]|S[table-format=1.1]}
\hline
\multirow{2}{*}{Differ In} & \multicolumn{2}{c|}{Baseline} & \multicolumn{2}{c}{Proposed} \\
\cline{2-5}
& {SI-SDRi} & {PESQ}  & {SI-SDRi} & {PESQ} \\
\hline
Random Cues      & 8.3 & 1.88  & 10.1 & 2.21 \\ \hline
All Cues          & 9.7  & 1.92  & 11.4 & 2.26 \\
Language          & 2.8 & 1.74  & 6.4 & 2.10 \\
Transcription     & -0.6 & 1.64  & 1.8 & 1.94  \\
Gender            & 8.9 & 1.90  & 11.0 & 2.25  \\
Emotion           & 2.0 & 1.55  & 1.9 & 1.77  \\
Pitch Level      & 5.3 & 1.77  & 6.0  & 2.05   \\
Pitch Range       & 2.9 & 1.61  & 3.0 & 1.96  \\
Loudness          & 7.1 & 1.81  & 8.1 & 2.11  \\
Distance          & 5.0 & 1.78  & 7.4 & 2.12  \\
Age Category     & -3.2 & 1.40 & -4.4 & 1.52  \\
Temporal Order    & 8.8 & 1.94 & 10.4 & 2.26 \\
Speaking Rate     & 0.7 & 1.69 & 1.5 & 1.93  \\
Speaking Duration & 6.4 & 1.87 & 6.9 & 2.15   \\
\hline
\end{tabular}
\end{table}


The results show the potential of training with only random cues, while using all cues leads to better performance by utilizing additional information during the inference stage. Notably, certain cues, such as gender and temporal order, exhibit strong discriminative power across different speech signals. These cues are language- and reverberation-independent and are more perceptible to humans than other individual cues.
Cues like speaking duration, pitch level, and loudness also perform well across languages and in reverberant environments. The pitch range cue doesn’t work as well as pitch level, likely because reverberation—especially under strong conditions—distorts the boundaries of pitch changes. In contrast, the age category performs poorly, likely due to limited training samples and its sensitivity to both language and reverberation effects. Additionally, some age-annotated speech samples exhibit emotion, further complicating differentiation. Meanwhile, transcription, emotion, and speaking rate cues outperform age category, as they are primarily influenced by language variations. 
The comparison between the baseline and proposed method indicates that integrating pre-trained WavLM Base+ features improves overall SI-SDRi, especially for language, transcription, gender, and distance cues, while also enhancing speech perception quality across all conditions.


\section{Conclusions and Future Work}
This study explored rich relative cues between speakers for text-guided target speech extraction. 
Experimental results suggest the effectiveness of incorporating multiple cues and leveraging pre-trained representations from WavLM Base+ CNN encoders. The findings also reveal that different cues have varying impacts and contributions in extracting target speakers' voices. 
Among the cues, gender and temporal order proved the most robust in cross-lingual and reverberant conditions, while cues like pitch level, loudness, distance, speaking duration, language, and pitch range also contributed, yielding over or nearly 3 dB SI-SDR gains for both models. Age category underperformed due to limited data and sensitivity to complex environment. Future work will expand age-labeled data, refine the model or training strategy to boost overall performance, and assess generalization across different languages and reverberation conditions.


\bibliographystyle{IEEEtran}
\bibliography{mybib}

\begin{thebibliography}{10}
\providecommand{\url}[1]{#1}
\csname url@samestyle\endcsname
\providecommand{\newblock}{\relax}
\providecommand{\bibinfo}[2]{#2}
\providecommand{\BIBentrySTDinterwordspacing}{\spaceskip=0pt\relax}
\providecommand{\BIBentryALTinterwordstretchfactor}{4}
\providecommand{\BIBentryALTinterwordspacing}{\spaceskip=\fontdimen2\font plus
\BIBentryALTinterwordstretchfactor\fontdimen3\font minus \fontdimen4\font\relax}
\providecommand{\BIBforeignlanguage}[2]{{%
\expandafter\ifx\csname l@#1\endcsname\relax
\typeout{** WARNING: IEEEtran.bst: No hyphenation pattern has been}%
\typeout{** loaded for the language `#1'. Using the pattern for}%
\typeout{** the default language instead.}%
\else
\language=\csname l@#1\endcsname
\fi
#2}}
\providecommand{\BIBdecl}{\relax}
\BIBdecl

\bibitem{Wang2019}
Q.~Wang, H.~Muckenhirn, K.~Wilson, P.~Sridhar, Z.~Wu, J.~R. Hershey, R.~A. Saurous, R.~J. Weiss, Y.~Jia, and I.~L. Moreno, ``{VoiceFilter: Targeted Voice Separation by Speaker-Conditioned Spectrogram Masking},'' in \emph{Proc. INTERSPEECH}, 2019, pp. 2728--2732.

\bibitem{vzmolikova2019speakerbeam}
K.~{\v{Z}}mol{\'\i}kov{\'a}, M.~Delcroix, K.~Kinoshita, T.~Ochiai, T.~Nakatani, L.~Burget, and J.~{\v{C}}ernock{\`y}, ``Speakerbeam: Speaker aware neural network for target speaker extraction in speech mixtures,'' \emph{IEEE Journal of Selected Topics in Signal Processing}, vol.~13, no.~4, pp. 800--814, 2019.

\bibitem{ge2022spex}
M.~Ge, C.~Xu, L.~Wang, E.~S. Chng, J.~Dang, and H.~Li, ``L-spex: Localized target speaker extraction,'' in \emph{IEEE International Conference on Acoustics, Speech and Signal Processing}, 2022, pp. 7287--7291.

\bibitem{ochiai2019multimodal}
T.~Ochiai, M.~Delcroix, K.~Kinoshita, A.~Ogawa, and T.~Nakatani, ``Multimodal speakerbeam: Single channel target speech extraction with audio-visual speaker clues.'' in \emph{INTERSPEECH}, 2019, pp. 2718--2722.

\bibitem{pan2022selective}
Z.~Pan, R.~Tao, C.~Xu, and H.~Li, ``Selective listening by synchronizing speech with lips,'' \emph{IEEE/ACM Transactions on Audio, Speech, and Language Processing}, vol.~30, pp. 1650--1664, 2022.

\bibitem{liu2024separate}
X.~Liu, Q.~Kong, Y.~Zhao, H.~Liu, Y.~Yuan, Y.~Liu, R.~Xia, Y.~Wang, M.~D. Plumbley, and W.~Wang, ``Separate anything you describe,'' \emph{IEEE/ACM Transactions on Audio, Speech, and Language Processing}, 2024.

\bibitem{li2023target}
C.~Li, Y.~Qian, Z.~Chen, D.~Wang, T.~Yoshioka, S.~Liu, Y.~Qian, and M.~Zeng, ``Target sound extraction with variable cross-modality clues,'' in \emph{IEEE International Conference on Acoustics, Speech and Signal Processing}, 2023, pp. 1--5.

\bibitem{jiang2024listen}
X.~Jiang, C.~Han, Y.~A. Li, and N.~Mesgarani, ``Listen, chat, and edit: Text-guided soundscape modification for enhanced auditory experience,'' \emph{arXiv preprint arXiv:2402.03710}, 2024.

\bibitem{hao2023typing}
X.~Hao, J.~Wu, J.~Yu, C.~Xu, and K.~C. Tan, ``Typing to listen at the cocktail party: Text-guided target speaker extraction,'' \emph{arXiv preprint arXiv:2310.07284}, 2023.

\bibitem{10890201}
M.~Huo, A.~Jain, C.~P. Huynh, F.~Kong, P.~Wang, Z.~Liu, and V.~Bhat, ``Beyond speaker identity: Text guided target speech extraction,'' in \emph{ICASSP 2025 - 2025 IEEE International Conference on Acoustics, Speech and Signal Processing}, 2025, pp. 1--5.

\bibitem{ji2024textrolspeech}
S.~Ji, J.~Zuo, M.~Fang, Z.~Jiang, F.~Chen, X.~Duan, B.~Huai, and Z.~Zhao, ``Textrolspeech: A text style control speech corpus with codec language text-to-speech models,'' in \emph{IEEE International Conference on Acoustics, Speech and Signal Processing}, 2024, pp. 10\,301--10\,305.

\bibitem{miller1981some}
J.~L. Miller, ``Some effects of speaking rate on phonetic perception,'' \emph{Phonetica}, vol.~38, no. 1-3, pp. 159--180, 1981.

\bibitem{quene2007just}
H.~Quen{\'e}, ``On the just noticeable difference for tempo in speech,'' \emph{Journal of Phonetics}, vol.~35, no.~3, pp. 353--362, 2007.

\bibitem{moore2012introduction}
B.~C. Moore, \emph{An Introduction to the Psychology of Hearing}.\hskip 1em plus 0.5em minus 0.4em\relax Brill, 2012.

\bibitem{blauert1997psychophysics}
J.~Blauert and S.~Hearing, ``The psychophysics of human sound localization,'' in \emph{Spatial Hearing}.\hskip 1em plus 0.5em minus 0.4em\relax MIT Press Cambridge, MA, USA, 1997.

\bibitem{linville2001vocal}
S.~Linville, \emph{Vocal Aging}.\hskip 1em plus 0.5em minus 0.4em\relax Singular Thomson Learning, 2001.

\bibitem{ryan1978age}
E.~B. Ryan and H.~L. Capadano~III, ``Age perceptions and evaluative reactions toward adult speakers,'' \emph{Journal of Gerontology}, vol.~33, no.~1, pp. 98--102, 1978.

\bibitem{zwicker2013psychoacoustics}
E.~Zwicker and H.~Fastl, \emph{Psychoacoustics: Facts and Models}.\hskip 1em plus 0.5em minus 0.4em\relax Springer Science \& Business Media, 2013, vol.~22.

\bibitem{zhang2008design}
J.~Zhang and H.~Jia, ``Design of speech corpus for mandarin text to speech,'' in \emph{The Blizzard Challenge 2008 Workshop}, 2008.

\bibitem{zhou2022emotional}
K.~Zhou, B.~Sisman, R.~Liu, and H.~Li, ``Emotional voice conversion: Theory, databases and esd,'' \emph{Speech Communication}, vol. 137, pp. 1--18, 2022.

\bibitem{shi21c_interspeech}
Y.~Shi, H.~Bu, X.~Xu, S.~Zhang, and M.~Li, ``Aishell-3: A multi-speaker mandarin tts corpus,'' in \emph{INTERSPEECH}, 2021, pp. 2756--2760.

\bibitem{magicdata2019}
{OpenSLR}, ``Magicdata mandarin chinese read speech corpus,'' \url{https://openslr.org/68/}, 2019, accessed: 2025-06-08.

\bibitem{panayotov2015librispeech}
V.~Panayotov, G.~Chen, D.~Povey, and S.~Khudanpur, ``{Librispeech: an asr corpus based on public domain audio books},'' in \emph{IEEE International Conference on Acoustics, Speech and Signal Processing}, 2015, pp. 5206--5210.

\bibitem{kerkeni2020french}
L.~Kerkeni, C.~Cleder, Y.~Serrestou, and K.~Raouf, ``French emotional speech database - oréau,'' Dec. 2020, {Zenodo}.

\bibitem{Pratap2020MLSAL}
V.~Pratap, Q.~Xu, A.~Sriram, G.~Synnaeve, and R.~Collobert, ``Mls: A large-scale multilingual dataset for speech research,'' \emph{ArXiv}, vol. abs/2012.03411, 2020.

\bibitem{burkhardt2005database}
F.~Burkhardt, A.~Paeschke, M.~Rolfes, W.~F. Sendlmeier, B.~Weiss \emph{et~al.}, ``A database of german emotional speech.'' in \emph{INTERSPEECH}, vol.~5, 2005, pp. 1517--1520.

\bibitem{garcia2024emomatchspanishdb}
E.~Garcia-Cuesta, A.~B. Salvador, and D.~G. P{\~a}ez, ``Emomatchspanishdb: study of speech emotion recognition machine learning models in a new spanish elicited database,'' \emph{Multimedia Tools and Applications}, vol.~83, no.~5, pp. 13\,093--13\,112, 2024.

\bibitem{diaz2021gpurir}
D.~Diaz-Guerra, A.~Miguel, and J.~R. Beltran, ``gpurir: A python library for room impulse response simulation with gpu acceleration,'' \emph{Multimedia Tools and Applications}, vol.~80, no.~4, pp. 5653--5671, 2021.

\bibitem{peng2024target}
J.~Peng, M.~Delcroix, T.~Ochiai, O.~Plchot, S.~Araki, and J.~{\v{C}}ernock{\`y}, ``Target speech extraction with pre-trained self-supervised learning models,'' in \emph{IEEE International Conference on Acoustics, Speech and Signal Processing}, 2024, pp. 10\,421--10\,425.

\bibitem{subakan2021attention}
C.~Subakan, M.~Ravanelli, S.~Cornell, M.~Bronzi, and J.~Zhong, ``Attention is all you need in speech separation,'' in \emph{ICASSP 2021-2021 IEEE International Conference on Acoustics, Speech and Signal Processing}, 2021, pp. 21--25.

\bibitem{le2019sdr}
J.~Le~Roux, S.~Wisdom, H.~Erdogan, and J.~R. Hershey, ``Sdr--half-baked or well done?'' in \emph{ICASSP 2019-2019 IEEE International Conference on Acoustics, Speech and Signal Processing}, 2019, pp. 626--630.

\bibitem{941023}
A.~Rix, J.~Beerends, M.~Hollier, and A.~Hekstra, ``Perceptual evaluation of speech quality ({PESQ})-a new method for speech quality assessment of telephone networks and codecs,'' in \emph{2001 IEEE International Conference on Acoustics, Speech, and Signal Processing}, vol.~2, 2001, pp. 749--752 vol.2.

\end{thebibliography}

\end{document}